\documentclass[superscriptaddress,prb,twocolumn,amsmath,amssymb]{revtex4}
\usepackage{graphicx,natbib}
\usepackage[caption=false]{subfig}

\newcommand{\RM }[1]{\mathrm{#1}}
\newcommand{\ave}[1]{ {\langle {#1} \rangle} }

\def\kB{ k_{\RM{B}} }

\def\Pc{\varphi}

\def\s2{{s_2}}
\def\Is2{{I_{\s2}}}

\def\drsq{{\delta r^2}}
\def\logdrsq{\log_{10}(\drsq)}
\def\Prsq{{P[\log_{10}(\drsq)]}}
\def\gT{{\widetilde{g}}}
\def\stwoT{{\widetilde{s}_2}}
\def\stwoTC{{\widetilde{s}_{2,\RM{C}}}}
\def\ntotT{{\widetilde{n}_{\RM{tot}}}}

\begin{document}

\title{Structural signatures of mobility on intermediate time scales
  in a supercooled fluid}

\author{William P. Krekelberg}
\affiliation{Department of Chemical
  Engineering, University of
  Texas at Austin, Austin, TX 78712.}  

\author{Venkat Ganesan}
\affiliation{Department of Chemical
  Engineering, University of
  Texas at Austin, Austin, TX 78712.}

\author{Thomas M. Truskett}
\email{truskett@che.utexas.edu}
\thanks{Corresponding Author} 
\affiliation{Department of Chemical
  Engineering, University of
  Texas at Austin, Austin, TX 78712.}  
 \affiliation{Institute for Theoretical Chemistry, 
   University of Texas at Austin, Austin, TX 78712.}

\begin{abstract}
We use computer simulations to explore the manner in which the particle 
displacements on
intermediate time scales in supercooled fluids correlate to their 
dynamic structural environment. The
fluid we study, a binary mixture of hard spheres, exhibits classic
signatures of dynamic heterogeneity, 
including a bifurcated single-particle displacement
distribution (i.e., subpopulations of immobile and mobile
particles).  We find that immobile particles, during the course of
their displacements, exhibit stronger average
pair correlations to their neighbors than mobile particles, but not necessarily
higher average coordination numbers.  We discuss how the correlation
between structure and single-particle dynamics depends on observation time. 

\end{abstract}
\maketitle

When fluids are supercooled (or overcompressed) toward their glass
transition, their single-particle dynamics undergo qualitative
changes.\cite{Perera1996Consequences-of, Kob1997Dynamical-Heter,
  Donati1998Stringlike-Coop, Yamamoto1998Dynamics-of-hig,
  Angell2000Relaxation-in-g, Ediger2000Spatially-heter,
  Schweizer2007Dynamical-fluct}
One example is the emergence of dynamic heterogeneity at
time scales intermediate between ballistic and 
diffusive regimes of particle motion, which manifests 
as spatially dependendent relaxation 
processes in the
liquid\cite{Kob1997Dynamical-Heter,Gebremichael2001Spatially-corre,Weeks2000Three-Dimension,Glotzer2000Time-dependent-,Vidal-Russell2000Direct-observat,Ediger2000Spatially-heter,Glotzer2000Spatially-heter,Lacevic2003Spatially-heter,Vogel2004Spatially-Heter,Andersen2005Molecular-dynam,Richert2002Heterogeneous-d,Kawasaki2007Correlation-bet}
and bifurcated (multi-peaked) probability distributions
associated with single-particle displacements.  The latter suggests the presence of
distinct subpopulations of particles with different mobilities on
these time scales.\cite{Yamamoto1998Heterogeneous-D,
  Puertas2005Dynamic-heterog, Reichman2005Comparison-of-D,
  Saltzman2006Non-Gaussian-ef, Schweizer2007Dynamical-fluct,
  Saltzman2008Large-amplitude}
Dynamic heterogeneities of this sort continue to attract wide interest because
of their perceived consequences for other
processes in deeply supercooled liquids including the breakdown of the
Stokes-Einstein relationship\cite{Stillinger1994Translation-rot,
  Tarjus1995Breakdown-of-th,
  Yamamoto1998Heterogeneous-D,
  Ediger2000Spatially-heter,
  Kumar2006Nature-of-the-b}
and the emergence of non-exponential trends in
relaxation of the structure
factor.\cite{Doliwa1999The-origin-of-a,Ediger2000Spatially-heter}

Several recent studies have systematically explored the extent to
which a fluid's particle
configuration at a given time influences the spatial distribution
of relaxation processes that immediately follow.\cite{Widmer-Cooper2004How-Reproducibl,
    Berthier2007Structure-and-d,Widmer-Cooper2005On-the-relation,Widmer-Cooper2006Predicting-the-,Widmer-Cooper2006Free-volume-can,Widmer-Cooper2007On-the-study-of}  Other investigations have focused on probing how certain
static structural properties of fluids can be used to correlate the 
effects that temperature, density, external fields, and interparticle
potential have on long-time dynamics.\cite{Rosenfeld1999A-quasi-univers,Dzugutov1996A-univeral-scal,Mittal2006Thermodynamics-,Goel2009Available-state,Krekelberg2009Generalized-,Gnan2009-}
Here, we explore a different, but related, question.  
Does the local structure surrounding particles in a deeply supercooled liquid, 
averaged over an intermediate
time scale relevant for dynamic heterogeneities, 
strongly correlate with their mean-square displacements during the
period of observation?
In other words, do particles with high mobility sample more
disordered structural environments during the course of their
displacements than those with low mobility 
and vice versa?     

We investigate this question via 
event-driven~\cite{Rapaport2004The-Art-of-Mole} molecular dynamics 
simulation of dense
binary fluid mixtures of hard spheres. We set the
ratio of particle diameters in these mixtures to
$\sigma_1/\sigma_2=1.3$ and the ratio of particle masses
to~$m_1/m_2=(\sigma_1/\sigma_2)^3$, parameters that mimic concentrated
colloidal suspensions that were recently investigated
experimentally.\cite{Nugent2007Colloidal-Glass} We simulate
$N_1=N_2=1000$ particles in a periodically-replicated cubic cell of
volume $V$.  We present results for particle packing fractions
$\Pc=\pi (N_1\sigma_1^3+N_2\sigma_2^3)/6V$ of $0.57$ and $0.582$,
which, as we show below, correspond to state points with unimodal and
bifurcated displacement distributions on intermediate time scales,
respectively.  For brevity, we report quantities that are implicitly
nondimensionalized by appropriate combinations of the length scale,
$l_{\RM{c}}=\sigma_2$ and time scale $t_{\RM{c}}=\sqrt{m_2
  \sigma_2^2/\kB T}$, where $\kB$ is Boltzmann's constant.  We focus
on the dynamics and structure of the smaller type 2 particles, but we
note that the behavior of the larger type 1 particles (not shown here)
is qualitatively similar, as might be expected given the mild
particle-size asymmetry of the fluid.

\begin{figure}[htbp]
  \centering
  \includegraphics[keepaspectratio,clip]{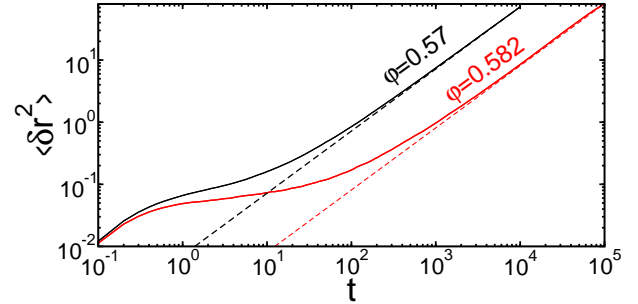}
\caption{(Color online) Mean-square displacement $\ave{\delta r^2}$
  of the smaller (type 2)
  particles versus time~$t$ at packing fraction $\Pc=0.57$ (black) and
  $0.582$ (red).  Dashed lines are fits of the Einstein relation $\ave{\drsq}=6Dt$ to the
  long-time behavior, resulting in tracer diffusion coefficients of 
  $D=1.2 \times 10^{-3}$ and $1.5
  \times 10^{-4}$ at $\Pc=0.57$ and $0.582$, respectively.}
  \label{fig:hetero:msd}
\end{figure}

We begin by examining the time~$t$ dependence of the average
mean-square displacement, $\ave{\drsq}$, for the type 2 particles.  In
particular, Figure~\ref{fig:hetero:msd} displays results for
packing fractions of $\Pc=0.57$ and $0.582$.  At both state points,
the fluid exhibits a mean-square displacement plateau at intermediate
times, which is characteristic of ``cage''
dynamics.\cite{Weeks2002Properties-of-C}  Schematically, the plateau
separates the ballistic motion that occurs at very short times (before
motion is temporarily hindered by collisions with the cage of
nearest-neighbor particles) and the diffusive motion that particles
ultimately exhibit at long times (after breaking through the cage).  As
should be expected, the plateau occurs at smaller displacements and
persists for longer times when $\Pc$ is increased, indicating that the
cage formed by the nearest-neighbor coordination shell becomes both
tighter and more difficult to disrupt at higher packing fraction.

\begin{figure}
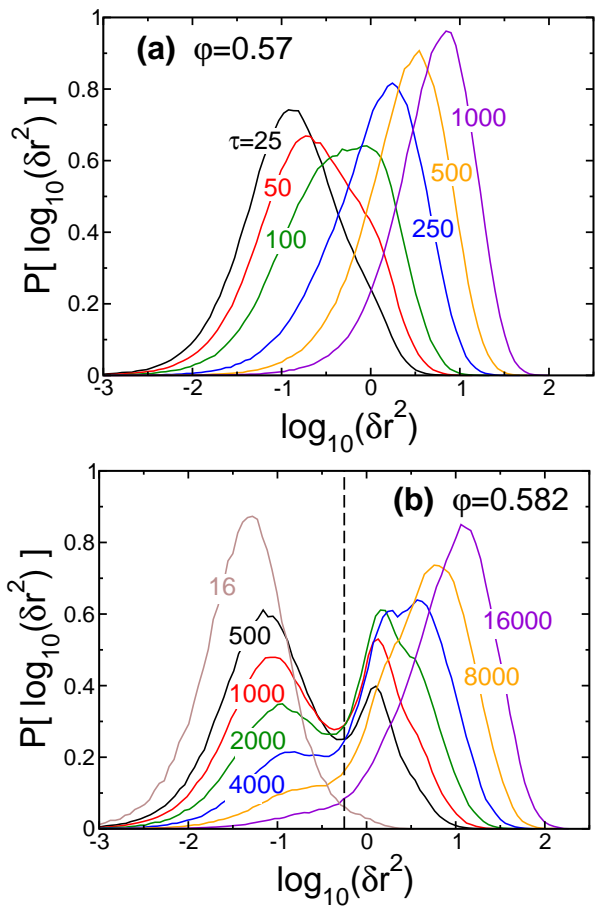

  \centering
  \includegraphics[keepaspectratio,clip]{logRsq_0.57_.eps}
  \subfloat{\label{fig:hetero:rsq_dist_057}}
  \includegraphics[keepaspectratio,clip]{logRsq_0.582_.eps}
  \subfloat{\label{fig:hetero:rsq_dist_058}}
  \caption{(Color online) Distribution of the logarithm of mean-square 
displacement
    $\Prsq$ for type 2 particles of the binary hard-sphere mixture
    described in the text.  Data cover several time intervals $t=\tau$ for
    packing fractions \textbf{(a)} $\Pc=0.57$ and \textbf{(b)}
    $\Pc=0.582$.  The numbers in the figure provide the value of
    $\tau$ for each curve.  In \textbf{(b)}, the vertical
    line is at $\drsq=0.56$, the approximate dividing line between
    mobile and immobile particles discussed in the text.}
  \label{fig:hetero:rsq_dist_all}
\end{figure}

In order to characterize dynamic heterogeneities of the type 2
particles at the state points examined above, we follow
Ref.~\onlinecite{Cates2004Theory-and-simu} and investigate the distributions
of the logarithm of the mean-square displacement ($\Prsq$) over a
variety of time intervals ($t=\tau$).  Specifically,
Figure~\ref{fig:hetero:rsq_dist_057} displays $\Prsq$ for the fluid at
$\Pc=0.57$ and values of $\tau$ that correspond to average mean-square
displacements which span from the plateau ``cage'' region to the
beginning of the diffusive regime (see Figure~\ref{fig:hetero:msd}).
The main point of Figure~\ref{fig:hetero:rsq_dist_057} is that, for
all values of $\tau$, $\Prsq$ remains unimodal, indicating that
pronounced single-particle dynamic heterogeneities have not yet
emerged in the fluid at this packing fraction.  However, also notice
that for the displacement distributions corresponding to intermediate
times, particularly at $\tau=50$, a shoulder at higher mean-square
displacements becomes evident.  This shoulder is a precursor to the
bifurcated displacement behavior that occurs at $\Pc=0.582$, which we
discuss in detail below.

Figure~\ref{fig:hetero:rsq_dist_058} displays the behavior of $\Prsq$
at $\Pc=0.582$.  Notice that, for intermediate times before diffusive
behavior is reached ($500\leq \tau \leq 4000$), $\Prsq$ shows
signatures of bimodality, i.e., two-peak structure.  
This suggests that subpopulations of
mobile and immobile particles are emerging. 
A qualitative dividing line between the two subpopulations
can be drawn at $\delta r^2\approx 0.56$ (vertical dashed line in
Figure~\ref{fig:hetero:msd}), which distinguishes the particles that
are still caged (i.e., localized) 
on intermediate time scales from those that have
broken through their nearest-neighbor coordination shells to attain
larger displacements.  For descriptive purposes, we label particles
with $\drsq<0.56$ ``immobile'' (on these time scales), while we label
those with $\drsq>0.56$ ``mobile''.  The type of dynamic behavior
depicted in Figure~\ref{fig:hetero:rsq_dist_058} has been well
documented in other systems.\cite{Donati1999Spatial-correla}  Below
we investigate whether the immobile particles experience, on average,
a more ordered structural environment than the mobile particles
during the time intervals of their respective displacements.

To carry out the analysis described above, we first examine our
simulation trajectories to accumulate statistics for each time
interval $\tau$, classifying type 2 particles according to the
logarithm of their mean-square displacement during the observation
period.  This amounts to creating a histogram from the distributions
shown in Figures~\ref{fig:hetero:rsq_dist_057} and
\ref{fig:hetero:rsq_dist_058}, assigning type 2 particles to
``mobility bins''.  Depending on the value of $\tau$, we use bin sizes
for $\logdrsq$
in the range 0.1 - 0.2, which we find is sufficiently
narrow to capture the shapes of the mobility distributions, but coarse
enough to allow for excellent statistical sampling.  Following
Ref.~\onlinecite{Palomar2008Study-of-spatia}, we compute an average
pair correlation function, $\gT_{2j}(r)$, between the type 2 particles in a
particular $\tau$-dependent mobility bin and {\em all} surrounding
particles of type $j$. In determining $\gT_{2j}(r)$, the relevant pair
separations are computed from configurations sampled uniformly in time 
throughout the time interval of length $\tau$.  
In this work, we use the $\sim$ overbar to denote a quantity that
describes an average dynamic structure surrounding the type 2
particles belonging to a specific
mobility bin for a given time interval $\tau$.

In order to convert the 
$\tau$-dependent structural information contained in the 
$\gT_{2j}(r)$ into a number that characterizes the degree of 
average pair translational order 
that a type 2 particle (in given a mobility bin) experiences
during the course of its displacement, we 
compute $-\stwoT$, 
which we define as
\begin{equation}
  \label{eq:hetero:s2_classi}
  -\stwoT \equiv \frac{\rho}{2} \sum_j x_j \int_0^{\infty} d{\bf r} \{
  \gT_{2j}({\bf r}) \ln
  \,\gT_{2j}({\bf r}) - [\gT_{2j}({\bf r})-1]\}.
\end{equation}
Here, $\rho=(N_1+N_2)/V$ is the total number density, and $x_j$ is the
mole fraction of component $j$.  This measure is a dynamic
generalization of a static structural metric, $-s_2^{(2)}$, which
quantifies the contribution to the excess entropy of a mixture arising
from equilibrium pair correlations involving particles of type $2$.
Our motivation for using $-\stwoT$ in this study comes from
(1) the earlier empirical
observation\cite{Samanta2001Universal-Scali,Pond2009Composition-and}
that the long-time tracer diffusivity of species $i$ in equilibrium
mixtures scales in a simple way with the static measure $-s_2^{(2)}$
and (2) the wider literature demonstrating that excess entropy
captures many of the effects that temperature, density, and
confinement have on the transport coefficients of equilibrium fluids
(see, e.g.,
Refs.~\onlinecite{Rosenfeld1999A-quasi-univers,Dzugutov1996A-univeral-scal,Mittal2006Thermodynamics-,Goel2009Available-state,Krekelberg2009Generalized-,Gnan2009-}).
Although we focus exclusively on the quantity $-\stwoT$ in this paper
to characterize dynamic structure, we have found that other commonly used
structural order metrics\cite{Truskett2000Towards-a-quant} calculable
from the $\tau$-averaged $\gT_{2j}(r)$ produce qualitatively similar results.

\begin{figure}[h]
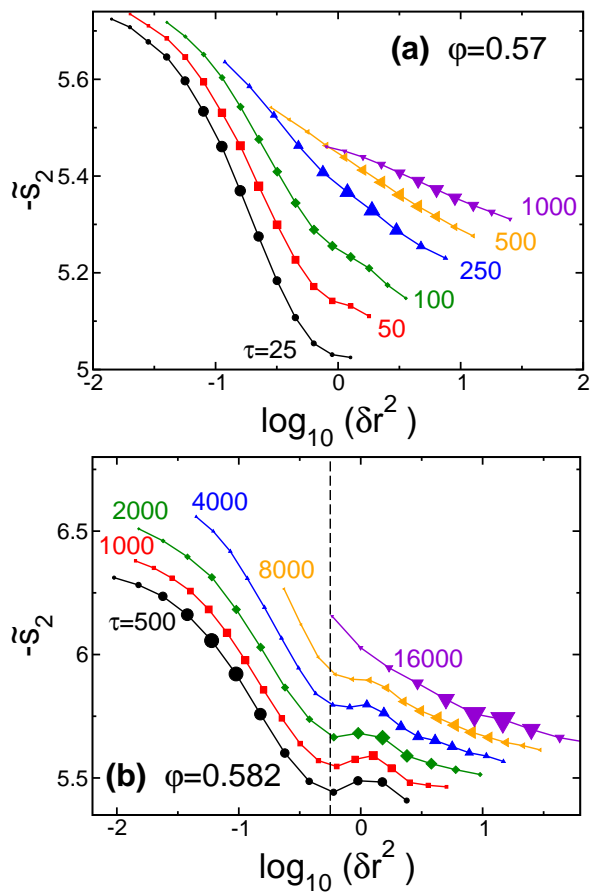

  \centering
  \includegraphics[keepaspectratio,clip]{s2_logr2_0.57_.eps}
  \subfloat{\label{fig:hetero:s2_vs_rsq_057}}
  \includegraphics[keepaspectratio,clip]{s2_logr2_0.582_.eps}
    \subfloat{\label{fig:hetero:s2_vs_rsq_058}}
    \caption{(Color online) Metric for the average structural order, $-\stwoT$,
      that type 2 particles of the binary mixture discussed in the
      text experience during an observation time $\tau$ as a function
      of the logarithm of their mean-square displacement
      $\log_{10}(\drsq)$.  Data is for packing fractions \textbf{(a)}
      $\Pc=0.57$ and \textbf{(b)} $\Pc=0.582$.  The numbers in the
      figure correspond to the value of $\tau$ for the curve of the
      same color.  The size of the symbol is proportional to the
      fraction of particles in the ``mobility bin'' centered at that
      value of $\log_{10}(\drsq)$.  In \textbf{(b)}, the horizontal
      dashed line at $\drsq=0.56$ represents the boundary between
      mobile and immobile particles (see
      Figure~\ref{fig:hetero:rsq_dist_058} and text).}
  \label{fig:hetero:s2_vs_rsq_all}
\end{figure}

To establish a baseline, we first examine the connection between displacement and average structural
order (during displacement) 
for type 2 particles in the system at $\Pc=0.57$, a supercooled
fluid state point for which pronounced dynamic heterogeneities have
not yet emerged.  Figure~\ref{fig:hetero:s2_vs_rsq_057} shows that,
for all times spanning from the plateau region to the diffusive limit of the
mean-square displacement curve in Figure~\ref{fig:hetero:msd},
there is a clear negative correlation between the average structural
order $-\stwoT$ surrounding a particle during an observation window
$\tau$ and how far it moves in that time frame.  
As might be expected,
the short-time curves are considerably steeper than longer-time
curves.    That is, for shorter time intervals, larger 
displacements are, on average, accompanied by progressively more
local structural disordering (i.e., weakening of the pair correlations
associated with the tagged particle).  At longer times approaching the
diffusive limit, particles sample a broader distribution containing 
larger mean-square displacements, and the correlation between average
structure and dynamics of the tagged particle is reduced.

\begin{figure}[!h]
  \centering
  \includegraphics[keepaspectratio,clip]{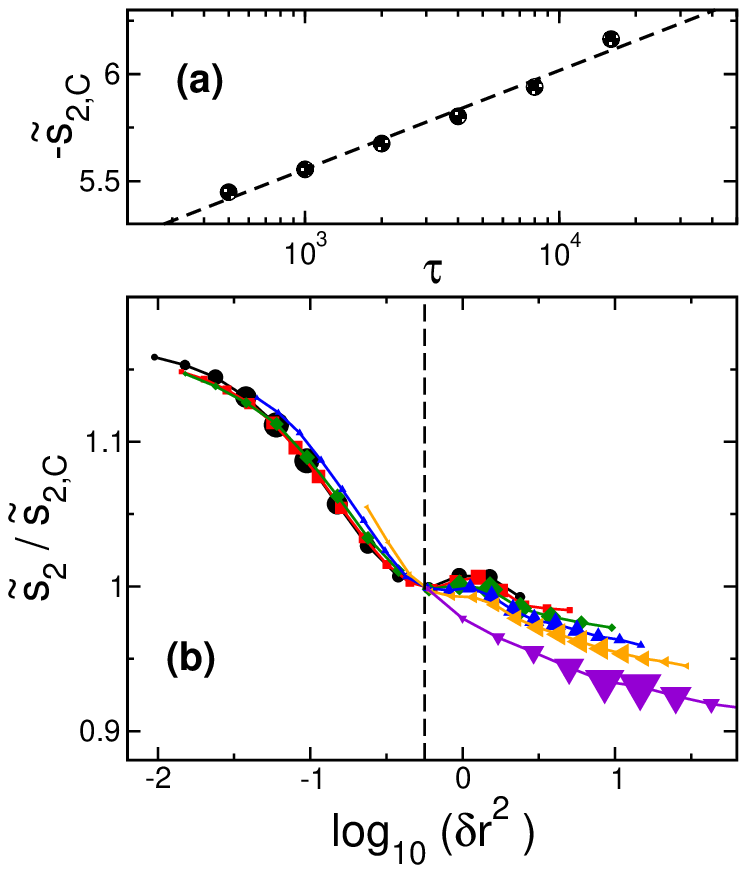}
  \caption{(Color online) (a)  The value of $\stwoT$ for $\drsq=0.56$,
      $\stwoTC=\stwoT(\drsq=0.56)$, as a function $\tau$.  (b) Reduced
      measure of structural order $\stwoT/\stwoTC(\tau)$ versus
      $\log_{10}(\drsq)$ for $\Pc=0.582$.  The data and symbols are the same as in 
      Figure~\ref{fig:hetero:s2_vs_rsq_058}.} 
    \label{fig:s2c}
\end{figure}

The behavior is different, however, for the fluid at $\Pc=0.582$ where
the bifurcated displacement distribution characteristic of strong dynamic
heterogeneities is observed (see
Figure~\ref{fig:hetero:s2_vs_rsq_058}).  In particular, there is now a
sharp change in the slope of the correlation between
average structure and dynamics when one compares ``immobile'' versus ``mobile" particles.  For immobile caged
particles, there is again a strong negative correlation between 
structural order and particle displacement; i.e.,
larger vibrational displacements are accompanied by increasingly 
weaker average pair
correlations with neighboring particles.  Moreover, the immobile
particles sample a structural environment that is, on average,
more ordered than mobile particles on the same time
scale. However, 
{\em within the class of mobile particles} at a given time, 
the correlation between structural order and mobility is considerably weaker.

In fact, if we focus on the structure and dynamics of 
mobile particles, a minor
secondary effect in Figure~\ref{fig:hetero:s2_vs_rsq_058} is also 
apparent.  Specifically, those particles with
intermediate displacements (of the order of a single particle
diameter) on a time scale $\tau$ can have slightly more average structural
order than those with either smaller or larger displacements.  This
feature is
likely due to recaging events,~\cite{Weeks2002Properties-of-C}
where a new coordination shell is temporarily formed around a particle
that has traveled just far enough to ``break free'' from its
original set of nearest neighbors.

Another prominent feature of Figure~\ref{fig:hetero:s2_vs_rsq_058} is
that the curves for different $\tau$ have the same generic 
shape and appear merely shifted in terms of their average structural order.
In fact, Fig~\ref{fig:s2c}(a) shows that the average structural order for
particles with displacements at the boundary between immobile and
mobile 
regions, i.e., $-\stwoTC\equiv-\stwoT(\drsq=0.56)$, increases
logarithmically with $\tau$ under these conditions.  This result
makes intuitive sense --- more structural order is, on average, expected to
surround particles that take a longer period of time to exhibit the
same value of mean-square displacement.  
Interestingly, as we show in  Fig~\ref{fig:s2c}(b), the $\tau$
dependence of Figure~\ref{fig:hetero:s2_vs_rsq_058} is approximately 
removed altogether if one simply plots the average structural order
of particles of a given mobility class 
normalized by its value at the boundary between mobile and immobile
particles, $\stwoT/\stwoTC$, versus $\log_{10}(\drsq)$.  In other
words, the overall ``scale'' of the structural
order surrounding particles has a simple $\tau$ dependence for the
intermediate times where dynamic heterogeneities are
present  ($500 \le \tau \le 8000$).  
The relative differences between the structures surrounding 
particles in different mobility classes, on the other hand, show only
a very weak dependence on $\tau$.  

Although we do not yet have a theoretical model for predicting 
the average pair correlations of
particles with different mobilities, the logarithmic time dependencies 
illustrated Fig.~\ref{fig:s2c} are perhaps not too surprising when one
considers the following.  Mean-square displacements of a tagged particle over a time
$\tau$ can be expressed $6D(\tau)\tau$.  Little is known about how
$D(\tau)$ relates to structure, but long-time 
self-diffusivities $D(\infty)$ 
exhibit an approximately exponential dependence on the two-body excess
entropy\cite{Dzugutov1996A-univeral-scal,Samanta2001Universal-Scali,Pond2009Composition-and}
for dense fluids.  Thus, to first approximation, one might 
expect that particles which undergo the same mean-square
displacement for different time scales $\tau$ share a constant value 
of the product $\tau \exp
[\stwoT]$.  Although this crude argument is consistent with the
data of Fig.~\ref{fig:s2c} and has some intuitive
appeal, it is far from rigorous.  More theoretical work will 
be needed to provide a comprehensive understanding of the
connections between dynamic structure and single-particle mobility 
observed here.

\begin{figure}
  \centering
  \includegraphics[keepaspectratio,clip]{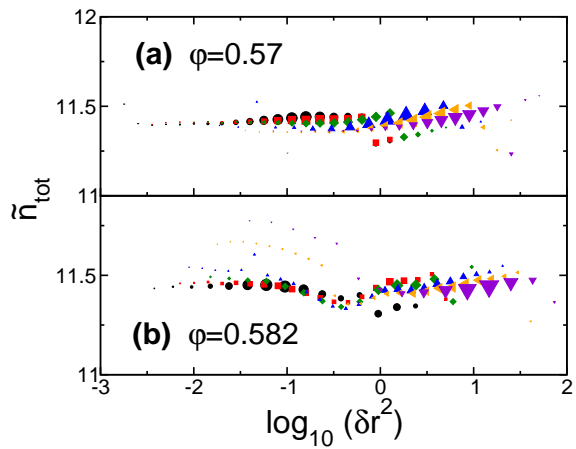}
  \subfloat{\label{fig:hetero:neighbors_panel_057}}
  \subfloat{\label{fig:hetero:neighbors_panel_0582}}

  \caption{(Color online) Average number of nearest neighbors,
    $\ntotT$, for type 2 particles of the binary mixture discussed in
    the text during an observation time $\tau$ as a function of the
    logarithm of their mean-square displacement $\log_{10}(\drsq)$.
    Data is for packing fractions \textbf{(a)} $\Pc=0.57$ and
    \textbf{(b)} $\Pc=0.582$. The symbols are the same as those in
    Figure~\ref{fig:hetero:s2_vs_rsq_all}.  The size of the symbols is
    proportional to the fraction of particles belonging to each
    mobility bin over time interval $\tau$.  }
  \label{fig:hetero:neighbors_panel_all}
\end{figure}

Given the above results, another question one might logically ask is 
whether a simpler
measure like coordination number, which roughly characterizes local density
surrounding a particle, might provide the same qualitative information
as $-\stwoT$.  To test this idea, we have also collected statistics on
 $\ntotT\equiv 4 \pi \sum_{j} x_j \rho \int_0^{r_{\RM{min},j}} r^2  \gT_{2j}(r) dr$, the average number of nearest
neighbors surrounding type 2 particles (in a given mobility bin on
time interval $\tau$).
Here, $r_{\RM{min},j}$ is the location of the
first minimum in $\gT_{2j}(r)$.  Figures~\ref{fig:hetero:neighbors_panel_057} and
\subref*{fig:hetero:neighbors_panel_0582} display $\ntotT$
as a function of $\logdrsq$ 
for $\Pc=0.57$ and $0.582$,
respectively.  In contrast to $-\stwoT$, 
the average coordination number during the displacement 
does not provide a clear indication of the different structural
environments that surround mobile versus immobile particles on
intermediate time scales.  

To summarize, we have presented computer simulation 
data on a model supercooled fluid showing a correlation between
average structure and mean-square displacements that is qualitatively
different for immobile and mobile particles over times relevant for
dynamic heterogeneities.  Interestingly, although the overall scale of the
structural order depends logarithmically on observation time, the relative
differences in structural order between mobile and immobile particles 
are largely independent of time.  We
provide a simple rationale for the aforementioned logarithmic time
dependence, but more theoretical work is needed to fully understand
these trends.  Finally, we show that the aforementioned 
structural differences between particles with different mobilities are 
not reflected by the coordination number.  

In addition to further work on this problem in the arena of molecular
simulation, it might be interesting to eventually explore 
whether the dynamical rules of kinetically
constrained lattice models for supercooled
liquids\cite{Ritort2003Glassy-dynamics}
produce behavior qualitatively consistent with that of Fig.~\ref{fig:hetero:s2_vs_rsq_all} from our molecular dynamics
simulations.  It might also be interesting to compare these results to
those from confocal microscopy experiments of dense colloidal
suspensions.\cite{Weeks2000Three-Dimension,Kegel2000Direct-Observat} 

We gratefully acknowledge helpful conversations with
Profs. K. S. Schweizer and P.L. Geissler about this work.
Two authors (T.M.T. and V.G.)
acknowledge support of the Welch Foundation (F-1696 and F-1599,
respectively).  One author (T.M.T) acknowledges financial support of
the National Science Foundation (CTS-0448721) and the David and Lucile
Packard Foundation.  Another author (V.G.) acknowledges support 
from the US Army Research Office under Grant No. W911NF-07-1-0268.  
The Texas Advanced Computing Center (TACC)
provided computational resources for this study.


\begin{thebibliography}{47}
\expandafter\ifx\csname natexlab\endcsname\relax\def\natexlab#1{#1}\fi
\expandafter\ifx\csname bibnamefont\endcsname\relax
  \def\bibnamefont#1{#1}\fi
\expandafter\ifx\csname bibfnamefont\endcsname\relax
  \def\bibfnamefont#1{#1}\fi
\expandafter\ifx\csname citenamefont\endcsname\relax
  \def\citenamefont#1{#1}\fi
\expandafter\ifx\csname url\endcsname\relax
  \def\url#1{\texttt{#1}}\fi
\expandafter\ifx\csname urlprefix\endcsname\relax\def\urlprefix{URL }\fi
\providecommand{\bibinfo}[2]{#2}
\providecommand{\eprint}[2][]{\url{#2}}

\bibitem[{\citenamefont{Perera and
  Harrowell}(1996)}]{Perera1996Consequences-of}
\bibinfo{author}{\bibfnamefont{D.~N.} \bibnamefont{Perera}} \bibnamefont{and}
  \bibinfo{author}{\bibfnamefont{P.}~\bibnamefont{Harrowell}},
  \bibinfo{journal}{Phys. Rev. E} \textbf{\bibinfo{volume}{54}},
  \bibinfo{pages}{1652} (\bibinfo{year}{1996}).

\bibitem[{\citenamefont{Kob et~al.}(1997)\citenamefont{Kob, Donati, Plimpton,
  Poole, and Glotzer}}]{Kob1997Dynamical-Heter}
\bibinfo{author}{\bibfnamefont{W.}~\bibnamefont{Kob}},
  \bibinfo{author}{\bibfnamefont{C.}~\bibnamefont{Donati}},
  \bibinfo{author}{\bibfnamefont{S.~J.} \bibnamefont{Plimpton}},
  \bibinfo{author}{\bibfnamefont{P.~H.} \bibnamefont{Poole}}, \bibnamefont{and}
  \bibinfo{author}{\bibfnamefont{S.~C.} \bibnamefont{Glotzer}},
  \bibinfo{journal}{Phys. Rev. Lett.} \textbf{\bibinfo{volume}{79}},
  \bibinfo{pages}{2827} (\bibinfo{year}{1997}).

\bibitem[{\citenamefont{Donati et~al.}(1998)\citenamefont{Donati, Douglas, Kob,
  Plimpton, Poole, and Glotzer}}]{Donati1998Stringlike-Coop}
\bibinfo{author}{\bibfnamefont{C.}~\bibnamefont{Donati}},
  \bibinfo{author}{\bibfnamefont{J.~F.} \bibnamefont{Douglas}},
  \bibinfo{author}{\bibfnamefont{W.}~\bibnamefont{Kob}},
  \bibinfo{author}{\bibfnamefont{S.~J.} \bibnamefont{Plimpton}},
  \bibinfo{author}{\bibfnamefont{P.~H.} \bibnamefont{Poole}}, \bibnamefont{and}
  \bibinfo{author}{\bibfnamefont{S.~C.} \bibnamefont{Glotzer}},
  \bibinfo{journal}{Phys. Rev. Lett.} \textbf{\bibinfo{volume}{80}},
  \bibinfo{pages}{2338} (\bibinfo{year}{1998}).

\bibitem[{\citenamefont{Yamamoto and
  Onuki}(1998{\natexlab{a}})}]{Yamamoto1998Dynamics-of-hig}
\bibinfo{author}{\bibfnamefont{R.}~\bibnamefont{Yamamoto}} \bibnamefont{and}
  \bibinfo{author}{\bibfnamefont{A.}~\bibnamefont{Onuki}},
  \bibinfo{journal}{Phys. Rev. E} \textbf{\bibinfo{volume}{58}},
  \bibinfo{pages}{3515} (\bibinfo{year}{1998}{\natexlab{a}}).

\bibitem[{\citenamefont{Angell et~al.}(2000)\citenamefont{Angell, Ngai,
  McKenna, McMillan, and Martin}}]{Angell2000Relaxation-in-g}
\bibinfo{author}{\bibfnamefont{C.~A.} \bibnamefont{Angell}},
  \bibinfo{author}{\bibfnamefont{K.~L.} \bibnamefont{Ngai}},
  \bibinfo{author}{\bibfnamefont{G.~B.} \bibnamefont{McKenna}},
  \bibinfo{author}{\bibfnamefont{P.~F.} \bibnamefont{McMillan}},
  \bibnamefont{and} \bibinfo{author}{\bibfnamefont{S.~W.}
  \bibnamefont{Martin}}, \bibinfo{journal}{J. Appl. Phys.}
  \textbf{\bibinfo{volume}{88}}, \bibinfo{pages}{3113} (\bibinfo{year}{2000}).

\bibitem[{\citenamefont{Ediger}(2000)}]{Ediger2000Spatially-heter}
\bibinfo{author}{\bibfnamefont{M.~D.} \bibnamefont{Ediger}},
  \bibinfo{journal}{Annu. Rev. Phys. Chem.} \textbf{\bibinfo{volume}{51}},
  \bibinfo{pages}{99} (\bibinfo{year}{2000}).

\bibitem[{\citenamefont{Schweizer}(2007)}]{Schweizer2007Dynamical-fluct}
\bibinfo{author}{\bibfnamefont{K.~S.} \bibnamefont{Schweizer}},
  \bibinfo{journal}{Curr. Opin. Colloid In.} \textbf{\bibinfo{volume}{12}},
  \bibinfo{pages}{297} (\bibinfo{year}{2007}).

\bibitem[{\citenamefont{Vidal~Russell and
  Israeloff}(2000)}]{Vidal-Russell2000Direct-observat}
\bibinfo{author}{\bibfnamefont{E.}~\bibnamefont{Vidal~Russell}}
  \bibnamefont{and} \bibinfo{author}{\bibfnamefont{N.~E.}
  \bibnamefont{Israeloff}}, \bibinfo{journal}{Nature}
  \textbf{\bibinfo{volume}{408}}, \bibinfo{pages}{695} (\bibinfo{year}{2000}).

\bibitem[{\citenamefont{Glotzer et~al.}(2000)\citenamefont{Glotzer, Novikov,
  and Schr{\o}der}}]{Glotzer2000Time-dependent-}
\bibinfo{author}{\bibfnamefont{S.~C.} \bibnamefont{Glotzer}},
  \bibinfo{author}{\bibfnamefont{V.~N.} \bibnamefont{Novikov}},
  \bibnamefont{and} \bibinfo{author}{\bibfnamefont{T.~B.}
  \bibnamefont{Schr{\o}der}}, \bibinfo{journal}{J. Chem. Phys.}
  \textbf{\bibinfo{volume}{112}}, \bibinfo{pages}{509} (\bibinfo{year}{2000}).

\bibitem[{\citenamefont{Glotzer}(2000)}]{Glotzer2000Spatially-heter}
\bibinfo{author}{\bibfnamefont{S.~C.} \bibnamefont{Glotzer}},
  \bibinfo{journal}{J. Non-Cryst. Solids} \textbf{\bibinfo{volume}{274}},
  \bibinfo{pages}{342} (\bibinfo{year}{2000}).

\bibitem[{\citenamefont{Weeks et~al.}(2000)\citenamefont{Weeks, Crocker,
  Levitt, Schofield, and Weitz}}]{Weeks2000Three-Dimension}
\bibinfo{author}{\bibfnamefont{E.~R.} \bibnamefont{Weeks}},
  \bibinfo{author}{\bibfnamefont{J.~C.} \bibnamefont{Crocker}},
  \bibinfo{author}{\bibfnamefont{A.~C.} \bibnamefont{Levitt}},
  \bibinfo{author}{\bibfnamefont{A.}~\bibnamefont{Schofield}},
  \bibnamefont{and} \bibinfo{author}{\bibfnamefont{D.~A.} \bibnamefont{Weitz}},
  \bibinfo{journal}{Science} \textbf{\bibinfo{volume}{287}},
  \bibinfo{pages}{627} (\bibinfo{year}{2000}).

\bibitem[{\citenamefont{Gebremichael et~al.}(2001)\citenamefont{Gebremichael,
  Schr{\o}der, Starr, and Glotzer}}]{Gebremichael2001Spatially-corre}
\bibinfo{author}{\bibfnamefont{Y.}~\bibnamefont{Gebremichael}},
  \bibinfo{author}{\bibfnamefont{T.~B.} \bibnamefont{Schr{\o}der}},
  \bibinfo{author}{\bibfnamefont{F.~W.} \bibnamefont{Starr}}, \bibnamefont{and}
  \bibinfo{author}{\bibfnamefont{S.~C.} \bibnamefont{Glotzer}},
  \bibinfo{journal}{Phys. Rev. E} \textbf{\bibinfo{volume}{64}},
  \bibinfo{pages}{051503} (\bibinfo{year}{2001}).

\bibitem[{\citenamefont{La{\v c}evi{\'c} et~al.}(2003)\citenamefont{La{\v
  c}evi{\'c}, Starr, Schr{\o}der, and Glotzer}}]{Lacevic2003Spatially-heter}
\bibinfo{author}{\bibfnamefont{N.}~\bibnamefont{La{\v c}evi{\'c}}},
  \bibinfo{author}{\bibfnamefont{F.~W.} \bibnamefont{Starr}},
  \bibinfo{author}{\bibfnamefont{T.~B.} \bibnamefont{Schr{\o}der}},
  \bibnamefont{and} \bibinfo{author}{\bibfnamefont{S.~C.}
  \bibnamefont{Glotzer}}, \bibinfo{journal}{J. Chem. Phys.}
  \textbf{\bibinfo{volume}{119}}, \bibinfo{pages}{7372} (\bibinfo{year}{2003}).

\bibitem[{\citenamefont{Vogel and Glotzer}(2004)}]{Vogel2004Spatially-Heter}
\bibinfo{author}{\bibfnamefont{M.}~\bibnamefont{Vogel}} \bibnamefont{and}
  \bibinfo{author}{\bibfnamefont{S.~C.} \bibnamefont{Glotzer}},
  \bibinfo{journal}{Phys. Rev. Lett.} \textbf{\bibinfo{volume}{92}},
  \bibinfo{pages}{255901} (\bibinfo{year}{2004}).

\bibitem[{\citenamefont{Kawasaki et~al.}(2007)\citenamefont{Kawasaki, Araki,
  and Tanaka}}]{Kawasaki2007Correlation-bet}
\bibinfo{author}{\bibfnamefont{T.}~\bibnamefont{Kawasaki}},
  \bibinfo{author}{\bibfnamefont{T.}~\bibnamefont{Araki}}, \bibnamefont{and}
  \bibinfo{author}{\bibfnamefont{H.}~\bibnamefont{Tanaka}},
  \bibinfo{journal}{Phys. Rev. Lett.} \textbf{\bibinfo{volume}{99}},
  \bibinfo{pages}{215701} (\bibinfo{year}{2007}).

\bibitem[{\citenamefont{Andersen}(2005)}]{Andersen2005Molecular-dynam}
\bibinfo{author}{\bibfnamefont{H.~C.} \bibnamefont{Andersen}},
  \bibinfo{journal}{Proceedings of the National Academy of Sciences of the
  United States of America} \textbf{\bibinfo{volume}{102}},
  \bibinfo{pages}{6686} (\bibinfo{year}{2005}).

\bibitem[{\citenamefont{Richert}(2002)}]{Richert2002Heterogeneous-d}
\bibinfo{author}{\bibfnamefont{R.}~\bibnamefont{Richert}}, \bibinfo{journal}{J.
  Phys.: Condens. Matter} \textbf{\bibinfo{volume}{14}}, \bibinfo{pages}{R703}
  (\bibinfo{year}{2002}).

\bibitem[{\citenamefont{Yamamoto and
  Onuki}(1998{\natexlab{b}})}]{Yamamoto1998Heterogeneous-D}
\bibinfo{author}{\bibfnamefont{R.}~\bibnamefont{Yamamoto}} \bibnamefont{and}
  \bibinfo{author}{\bibfnamefont{A.}~\bibnamefont{Onuki}},
  \bibinfo{journal}{Phys. Rev. Lett.} \textbf{\bibinfo{volume}{81}},
  \bibinfo{pages}{4915} (\bibinfo{year}{1998}{\natexlab{b}}).

\bibitem[{\citenamefont{Puertas et~al.}(2005)\citenamefont{Puertas, Fuchs, and
  Cates}}]{Puertas2005Dynamic-heterog}
\bibinfo{author}{\bibfnamefont{A.~M.} \bibnamefont{Puertas}},
  \bibinfo{author}{\bibfnamefont{M.}~\bibnamefont{Fuchs}}, \bibnamefont{and}
  \bibinfo{author}{\bibfnamefont{M.~E.} \bibnamefont{Cates}},
  \bibinfo{journal}{cond-mat} \textbf{\bibinfo{volume}{1}},
  \bibinfo{pages}{0510443} (\bibinfo{year}{2005}).

\bibitem[{\citenamefont{Reichman et~al.}(2005)\citenamefont{Reichman, Rabani,
  and Geissler}}]{Reichman2005Comparison-of-D}
\bibinfo{author}{\bibfnamefont{D.~R.} \bibnamefont{Reichman}},
  \bibinfo{author}{\bibfnamefont{E.}~\bibnamefont{Rabani}}, \bibnamefont{and}
  \bibinfo{author}{\bibfnamefont{P.~L.} \bibnamefont{Geissler}},
  \bibinfo{journal}{J. Phys. Chem. B} \textbf{\bibinfo{volume}{109}},
  \bibinfo{pages}{14654} (\bibinfo{year}{2005}).

\bibitem[{\citenamefont{Saltzman and
  Schweizer}(2006)}]{Saltzman2006Non-Gaussian-ef}
\bibinfo{author}{\bibfnamefont{E.~J.} \bibnamefont{Saltzman}} \bibnamefont{and}
  \bibinfo{author}{\bibfnamefont{K.~S.} \bibnamefont{Schweizer}},
  \bibinfo{journal}{Phys. Rev. E} \textbf{\bibinfo{volume}{74}},
  \bibinfo{pages}{061501} (\bibinfo{year}{2006}).

\bibitem[{\citenamefont{Saltzman and
  Schweizer}(2008)}]{Saltzman2008Large-amplitude}
\bibinfo{author}{\bibfnamefont{E.~J.} \bibnamefont{Saltzman}} \bibnamefont{and}
  \bibinfo{author}{\bibfnamefont{K.~S.} \bibnamefont{Schweizer}},
  \bibinfo{journal}{Phys. Rev. E} \textbf{\bibinfo{volume}{77}},
  \bibinfo{pages}{051504} (\bibinfo{year}{2008}).

\bibitem[{\citenamefont{Stillinger and
  Hodgdon}(1994)}]{Stillinger1994Translation-rot}
\bibinfo{author}{\bibfnamefont{F.~H.} \bibnamefont{Stillinger}}
  \bibnamefont{and} \bibinfo{author}{\bibfnamefont{J.~A.}
  \bibnamefont{Hodgdon}}, \bibinfo{journal}{Phys. Rev. E}
  \textbf{\bibinfo{volume}{50}}, \bibinfo{pages}{2064} (\bibinfo{year}{1994}).

\bibitem[{\citenamefont{Tarjus and Kivelson}(1995)}]{Tarjus1995Breakdown-of-th}
\bibinfo{author}{\bibfnamefont{G.}~\bibnamefont{Tarjus}} \bibnamefont{and}
  \bibinfo{author}{\bibfnamefont{D.}~\bibnamefont{Kivelson}},
  \bibinfo{journal}{J. Chem. Phys.} \textbf{\bibinfo{volume}{103}},
  \bibinfo{pages}{3071} (\bibinfo{year}{1995}).

\bibitem[{\citenamefont{Kumar et~al.}(2006)\citenamefont{Kumar, Szamel, and
  Douglas}}]{Kumar2006Nature-of-the-b}
\bibinfo{author}{\bibfnamefont{S.~K.} \bibnamefont{Kumar}},
  \bibinfo{author}{\bibfnamefont{G.}~\bibnamefont{Szamel}}, \bibnamefont{and}
  \bibinfo{author}{\bibfnamefont{J.~F.} \bibnamefont{Douglas}},
  \bibinfo{journal}{J. Chem. Phys.} \textbf{\bibinfo{volume}{124}},
  \bibinfo{pages}{214501} (\bibinfo{year}{2006}).

\bibitem[{\citenamefont{Doliwa and Heuer}(1999)}]{Doliwa1999The-origin-of-a}
\bibinfo{author}{\bibfnamefont{B.}~\bibnamefont{Doliwa}} \bibnamefont{and}
  \bibinfo{author}{\bibfnamefont{A.}~\bibnamefont{Heuer}}, \bibinfo{journal}{J.
  Phys.: Condens. Matter} \textbf{\bibinfo{volume}{11}}, \bibinfo{pages}{A277}
  (\bibinfo{year}{1999}).

\bibitem[{\citenamefont{Widmer-Cooper et~al.}(2004)\citenamefont{Widmer-Cooper,
  Harrowell, and Fynewever}}]{Widmer-Cooper2004How-Reproducibl}
\bibinfo{author}{\bibfnamefont{A.}~\bibnamefont{Widmer-Cooper}},
  \bibinfo{author}{\bibfnamefont{P.}~\bibnamefont{Harrowell}},
  \bibnamefont{and}
  \bibinfo{author}{\bibfnamefont{H.}~\bibnamefont{Fynewever}},
  \bibinfo{journal}{Phys. Rev. Lett.} \textbf{\bibinfo{volume}{93}},
  \bibinfo{pages}{135701} (\bibinfo{year}{2004}).

\bibitem[{\citenamefont{Berthier and Jack}(2007)}]{Berthier2007Structure-and-d}
\bibinfo{author}{\bibfnamefont{L.}~\bibnamefont{Berthier}} \bibnamefont{and}
  \bibinfo{author}{\bibfnamefont{R.~L.} \bibnamefont{Jack}},
  \bibinfo{journal}{Phys. Rev. E} \textbf{\bibinfo{volume}{76}},
  \bibinfo{pages}{041509} (\bibinfo{year}{2007}).

\bibitem[{\citenamefont{Widmer-Cooper and
  Harrowell}(2005)}]{Widmer-Cooper2005On-the-relation}
\bibinfo{author}{\bibfnamefont{A.}~\bibnamefont{Widmer-Cooper}}
  \bibnamefont{and}
  \bibinfo{author}{\bibfnamefont{P.}~\bibnamefont{Harrowell}},
  \bibinfo{journal}{J. Phys.: Condens. Matter} \textbf{\bibinfo{volume}{17}},
  \bibinfo{pages}{S4025} (\bibinfo{year}{2005}).

\bibitem[{\citenamefont{Widmer-Cooper and
  Harrowell}(2006{\natexlab{a}})}]{Widmer-Cooper2006Predicting-the-}
\bibinfo{author}{\bibfnamefont{A.}~\bibnamefont{Widmer-Cooper}}
  \bibnamefont{and}
  \bibinfo{author}{\bibfnamefont{P.}~\bibnamefont{Harrowell}},
  \bibinfo{journal}{Phys. Rev. Lett.} \textbf{\bibinfo{volume}{96}},
  \bibinfo{pages}{185701} (\bibinfo{year}{2006}{\natexlab{a}}).

\bibitem[{\citenamefont{Widmer-Cooper and
  Harrowell}(2006{\natexlab{b}})}]{Widmer-Cooper2006Free-volume-can}
\bibinfo{author}{\bibfnamefont{A.}~\bibnamefont{Widmer-Cooper}}
  \bibnamefont{and}
  \bibinfo{author}{\bibfnamefont{P.}~\bibnamefont{Harrowell}},
  \bibinfo{journal}{J. Non-Cryst. Solids} \textbf{\bibinfo{volume}{352}},
  \bibinfo{pages}{5098} (\bibinfo{year}{2006}{\natexlab{b}}).

\bibitem[{\citenamefont{Widmer-Cooper and
  Harrowell}(2007)}]{Widmer-Cooper2007On-the-study-of}
\bibinfo{author}{\bibfnamefont{A.}~\bibnamefont{Widmer-Cooper}}
  \bibnamefont{and}
  \bibinfo{author}{\bibfnamefont{P.}~\bibnamefont{Harrowell}},
  \bibinfo{journal}{J. Chem. Phys.} \textbf{\bibinfo{volume}{126}},
  \bibinfo{pages}{154503} (\bibinfo{year}{2007}).


\bibitem[{\citenamefont{Rosenfeld}(1999)}]{Rosenfeld1999A-quasi-univers}
\bibinfo{author}{\bibfnamefont{Y.}~\bibnamefont{Rosenfeld}},
  \bibinfo{journal}{J. Phys.: Condens. Matter} \textbf{\bibinfo{volume}{11}},
  \bibinfo{pages}{5415} (\bibinfo{year}{1999}).


\bibitem[{\citenamefont{Dzugutov}(1996)}]{Dzugutov1996A-univeral-scal}
\bibinfo{author}{\bibfnamefont{M.}~\bibnamefont{Dzugutov}},
  \bibinfo{journal}{Nature} \textbf{\bibinfo{volume}{381}},
  \bibinfo{pages}{137} (\bibinfo{year}{1996}).

\bibitem[{\citenamefont{Mittal et~al.}(2006)\citenamefont{Mittal, Errington,
  and Truskett}}]{Mittal2006Thermodynamics-}
\bibinfo{author}{\bibfnamefont{J.}~\bibnamefont{Mittal}},
  \bibinfo{author}{\bibfnamefont{J.~R.} \bibnamefont{Errington}},
  \bibnamefont{and} \bibinfo{author}{\bibfnamefont{T.~M.}
  \bibnamefont{Truskett}}, \bibinfo{journal}{Phys. Rev. Lett.}
  \textbf{\bibinfo{volume}{96}}, \bibinfo{pages}{177804}
  (\bibinfo{year}{2006}).


\bibitem[{\citenamefont{Goel et~al.}(2009)\citenamefont{Goel, Krekelberg, Pond,
  Mittal, Shen, Errington, and Truskett}}]{Goel2009Available-state}
\bibinfo{author}{\bibfnamefont{G.}~\bibnamefont{Goel}},
  \bibinfo{author}{\bibfnamefont{W.~P.} \bibnamefont{Krekelberg}},
  \bibinfo{author}{\bibfnamefont{M.~J.} \bibnamefont{Pond}},
  \bibinfo{author}{\bibfnamefont{J.}~\bibnamefont{Mittal}},
  \bibinfo{author}{\bibfnamefont{V.~K.} \bibnamefont{Shen}},
  \bibinfo{author}{\bibfnamefont{J.~R.} \bibnamefont{Errington}},
  \bibnamefont{and} \bibinfo{author}{\bibfnamefont{T.~M.}
  \bibnamefont{Truskett}}, \bibinfo{journal}{J. Stat. Mech.}
  \textbf{\bibinfo{volume}{2009}}, \bibinfo{pages}{P04006}
  (\bibinfo{year}{2009}).


\bibitem[{\citenamefont{Krekelberg et~al.}(2009)\citenamefont{Krekelberg, Pond,
  Goel, Shen, Errington, and Truskett}}]{Krekelberg2009Generalized-}
  \bibinfo{author}{\bibfnamefont{W.~P.} \bibnamefont{Krekelberg}},
  \bibinfo{author}{\bibfnamefont{M.~J.} \bibnamefont{Pond}},
\bibinfo{author}{\bibfnamefont{G.}~\bibnamefont{Goel}},
  \bibinfo{author}{\bibfnamefont{V.~K.} \bibnamefont{Shen}},
  \bibinfo{author}{\bibfnamefont{J.~R.} \bibnamefont{Errington}},
  \bibnamefont{and} \bibinfo{author}{\bibfnamefont{T.~M.}
  \bibnamefont{Truskett}}, \bibinfo{journal}{Phys. Rev. E}
  \textbf{\bibinfo{volume}{80}}, \bibinfo{pages}{061205}
  (\bibinfo{year}{2009}).


\bibitem[{\citenamefont{Gnan et~al.}(2006)\citenamefont{Gnan,
      Schroder, Pedersen, Bailey, and Dyre}}]{Gnan2009-}
\bibinfo{author}{\bibfnamefont{N.}~\bibnamefont{Gnan}},
  \bibinfo{author}{\bibfnamefont{T.~B.} \bibnamefont{Schroder}}, \bibinfo{author}{\bibfnamefont{U.~R.}
  \bibnamefont{Pedersen}}, \bibinfo{author}{\bibfnamefont{N.~P.}
  \bibnamefont{Bailey}}, \bibnamefont{and}
\bibinfo{author}\bibnamefont{J.~P.} \bibnamefont{Dyre},
\bibinfo{journal}{J. Chem. Phys.}
  \textbf{\bibinfo{volume}{131}}, \bibinfo{pages}{234504}
  (\bibinfo{year}{2009}).


\bibitem[{\citenamefont{Kegel et~al.}(2000)\citenamefont{Kegel, , and van
  Blaaderen}}]{Kegel2000Direct-Observat}
\bibinfo{author}{\bibfnamefont{W.~K.} \bibnamefont{Kegel}} \bibnamefont{and}
  \bibinfo{author}{\bibfnamefont{A.}~\bibnamefont{van Blaaderen}},
  \bibinfo{journal}{Science} \textbf{\bibinfo{volume}{287}},
  \bibinfo{pages}{290} (\bibinfo{year}{2000}).


\bibitem[{\citenamefont{Ritort and Sollich}(2003)}]{Ritort2003Glassy-dynamics}
\bibinfo{author}{\bibfnamefont{F.}~\bibnamefont{Ritort}} \bibnamefont{and}
  \bibinfo{author}{\bibfnamefont{P.}~\bibnamefont{Sollich}},
  \bibinfo{journal}{Adv. Phys.} \textbf{\bibinfo{volume}{52}},
  \bibinfo{pages}{219} (\bibinfo{year}{2003}).



\bibitem[{\citenamefont{Rapaport}(2004)}]{Rapaport2004The-Art-of-Mole}
\bibinfo{author}{\bibfnamefont{D.~C.} \bibnamefont{Rapaport}},
  \emph{\bibinfo{title}{The Art of Molecular Dynamic Simulation}}
  (\bibinfo{publisher}{Cambridge University Press, Cambridge},
  \bibinfo{year}{2004}), \bibinfo{edition}{2nd} ed.

\bibitem[{\citenamefont{Nugent et~al.}(2007)\citenamefont{Nugent, Edmond,
  Patel, and Weeks}}]{Nugent2007Colloidal-Glass}
\bibinfo{author}{\bibfnamefont{C.~R.} \bibnamefont{Nugent}},
  \bibinfo{author}{\bibfnamefont{K.~V.} \bibnamefont{Edmond}},
  \bibinfo{author}{\bibfnamefont{H.~N.} \bibnamefont{Patel}}, \bibnamefont{and}
  \bibinfo{author}{\bibfnamefont{E.~R.} \bibnamefont{Weeks}},
  \bibinfo{journal}{Phys. Rev. Lett.} \textbf{\bibinfo{volume}{99}},
  \bibinfo{pages}{025702} (\bibinfo{year}{2007}).

\bibitem[{\citenamefont{Weeks and Weitz}(2002)}]{Weeks2002Properties-of-C}
\bibinfo{author}{\bibfnamefont{E.~R.} \bibnamefont{Weeks}} \bibnamefont{and}
  \bibinfo{author}{\bibfnamefont{D.~A.} \bibnamefont{Weitz}},
  \bibinfo{journal}{Phys. Rev. Lett.} \textbf{\bibinfo{volume}{89}},
  \bibinfo{pages}{095704} (\bibinfo{year}{2002}).

\bibitem[{\citenamefont{Cates et~al.}(2004)\citenamefont{Cates, Fuchs, Kroy,
  Poon, and Puertas}}]{Cates2004Theory-and-simu}
\bibinfo{author}{\bibfnamefont{M.~E.} \bibnamefont{Cates}},
  \bibinfo{author}{\bibfnamefont{M.}~\bibnamefont{Fuchs}},
  \bibinfo{author}{\bibfnamefont{K.}~\bibnamefont{Kroy}},
  \bibinfo{author}{\bibfnamefont{W.~C.~K.} \bibnamefont{Poon}},
  \bibnamefont{and} \bibinfo{author}{\bibfnamefont{A.~M.}
  \bibnamefont{Puertas}}, \bibinfo{journal}{J. Phys.: Condens. Matter}
  \textbf{\bibinfo{volume}{16}}, \bibinfo{pages}{S4861} (\bibinfo{year}{2004}).

\bibitem[{\citenamefont{Donati et~al.}(1999)\citenamefont{Donati, Glotzer,
  Poole, Kob, and Plimpton}}]{Donati1999Spatial-correla}
\bibinfo{author}{\bibfnamefont{C.}~\bibnamefont{Donati}},
  \bibinfo{author}{\bibfnamefont{S.~C.} \bibnamefont{Glotzer}},
  \bibinfo{author}{\bibfnamefont{P.~H.} \bibnamefont{Poole}},
  \bibinfo{author}{\bibfnamefont{W.}~\bibnamefont{Kob}}, \bibnamefont{and}
  \bibinfo{author}{\bibfnamefont{S.~J.} \bibnamefont{Plimpton}},
  \bibinfo{journal}{Phys. Rev. E} \textbf{\bibinfo{volume}{60}},
  \bibinfo{pages}{3107} (\bibinfo{year}{1999}).

\bibitem[{\citenamefont{Palomar and
  Ses\'{e}}(2008)}]{Palomar2008Study-of-spatia}
\bibinfo{author}{\bibfnamefont{R.}~\bibnamefont{Palomar}} \bibnamefont{and}
  \bibinfo{author}{\bibfnamefont{G.}~\bibnamefont{Ses\'{e}}},
  \bibinfo{journal}{J. Chem. Phys.} \textbf{\bibinfo{volume}{129}},
  \bibinfo{pages}{064505} (\bibinfo{year}{2008}).

\bibitem[{\citenamefont{Samanta et~al.}(2001)\citenamefont{Samanta, Ali, and
  Ghosh}}]{Samanta2001Universal-Scali}
\bibinfo{author}{\bibfnamefont{A.}~\bibnamefont{Samanta}},
  \bibinfo{author}{\bibfnamefont{S.~M.} \bibnamefont{Ali}}, \bibnamefont{and}
  \bibinfo{author}{\bibfnamefont{S.~K.} \bibnamefont{Ghosh}},
  \bibinfo{journal}{Phys. Rev. Lett.} \textbf{\bibinfo{volume}{87}},
  \bibinfo{pages}{245901} (\bibinfo{year}{2001}).

\bibitem[{\citenamefont{Pond et~al.}(2009)\citenamefont{Pond, Krekelberg, Shen,
  Errington, and Truskett}}]{Pond2009Composition-and}
\bibinfo{author}{\bibfnamefont{M.~J.} \bibnamefont{Pond}},
  \bibinfo{author}{\bibfnamefont{W.~P.} \bibnamefont{Krekelberg}},
  \bibinfo{author}{\bibfnamefont{V.~K.} \bibnamefont{Shen}},
  \bibinfo{author}{\bibfnamefont{J.~R.} \bibnamefont{Errington}},
  \bibnamefont{and} \bibinfo{author}{\bibfnamefont{T.~M.}
  \bibnamefont{Truskett}}, \bibinfo{journal}{J. Chem. Phys.}
  \textbf{\bibinfo{volume}{131}}, \bibinfo{pages}{161101}
  (\bibinfo{year}{2009}).


\bibitem[{\citenamefont{Truskett et~al.}(2000)\citenamefont{Truskett, Torquato,
  and Debenedetti}}]{Truskett2000Towards-a-quant}
\bibinfo{author}{\bibfnamefont{T.~M.} \bibnamefont{Truskett}},
  \bibinfo{author}{\bibfnamefont{S.}~\bibnamefont{Torquato}}, \bibnamefont{and}
  \bibinfo{author}{\bibfnamefont{P.~G.} \bibnamefont{Debenedetti}},
  \bibinfo{journal}{Phys. Rev. E} \textbf{\bibinfo{volume}{62}},
  \bibinfo{pages}{993} (\bibinfo{year}{2000}).

\end{thebibliography}
\end{document}